\documentclass[%
superscriptaddress,
showpacs,
amsmath,amssymb,
aps,
prc,
showkeys,
twocolumn,
]{revtex4-1}

\usepackage{amsmath,mathrsfs,amssymb}
\usepackage{eurosym}
\usepackage{graphicx,subfigure}
\usepackage{color}
\usepackage{indentfirst}
\usepackage{extarrows}
\usepackage{hyperref}
\hypersetup{colorlinks=true, citecolor=blue, urlcolor=blue, linkcolor=blue}

\def\nuc#1#2{\relax\ifmmode{}^{#1}{\protect\text{#2}}\else${}^{#1}$#2\fi}

\begin{document}
    
\title{Possible determination of high-lying single particle components with $(p,d)$ reactions}
    
\author{Y.P.~Xu}
\affiliation{School of Physics and Nuclear Energy Engineering, Beihang University,
    Beijing 100191, China}

\author{D.Y. Pang}
\email[Corresponding author: ]{dypang@buaa.edu.cn}
\affiliation{School of Physics and Nuclear Energy Engineering, Beihang University, Beijing 100191, China}
\affiliation{Beijing Key Laboratory of Advanced Nuclear Materials and Physics, Beihang University, Beijing 100191, China}

\author{X.Y. Yun}
\affiliation{School of Physics and Nuclear Energy Engineering, Beihang University,
    Beijing 100191, China}

\author{S. Kubono}
\affiliation{RIKEN Nishina Center, 2-1 Hirosawa, Wako, Saitama 351-0198, Japan}
\affiliation{Center for Nuclear Study, the University of Tokyo, 2-1 Hirosawa, Wako, Saitama 351-0198, Japan}

\author{C.A.~Bertulani}
\affiliation{Texas A\&M University-Commerce, 75428, Commerce, TX, United States of America}

\author{C.X.~Yuan}
\affiliation{Sino-French Institute of Nuclear Engineering and Technology, Sun Yat-Sen University, Zhuhai 519082, China}
    
    \begin{abstract}
        A detailed feasibility study on deducing the high-lying single-particle components (HLSPCs), which are important but used to be ignored, in the ground and low-lying excited states of even-even light nuclei is performed by analyses of $(p,d)$ reactions with \nuc{12}{C}, \nuc{24}{Mg}, \nuc{28}{Si}, and \nuc{40}{Ca} targets at 51.93 MeV. Coupled reaction channels (CRC) analyses have been made for $(p,d)$ transitions to the $j$-forbidden excited states in \nuc{11}{C} (${\tfrac{5}{2}}^-$, 4.32 MeV), \nuc{23}{Mg} (${\tfrac{7}{2}}^+$, 2.05 MeV), \nuc{27}{Si}  (${\tfrac{7}{2}}^+$, 2.16 MeV) and \nuc{39}{Ca} (${\tfrac{9}{2}}^-$, 3.64 MeV), including the major allowed transition components together with direct components of HLSPCs. Spectroscopic amplitudes of the HLSPCs are deduced by fitting the angular distributions of the ground and the $j$-forbidden excited states simultaneously. The present analysis demonstrates for the first time that information about HLSPCs in atomic nuclei can be obtained from analysis of $(p,d)$ reactions.
    \end{abstract}
    
\maketitle

\section{Introduction}

Admixture of high-lying single-particle components (HLSPCs) has been found to be important for the structure of atomic nuclei for a long time \cite{Hyuga-NPA-1980, Warburton-PLB-1992}. In modern \textit{ab initio} calculations, use of very large basis space, such as 10 $\hbar\omega$, are also found to be necessary for reasonable reproduction of the energy levels of light nuclei \cite{Barrett-PPNP-2013, Calci-PRL-2016}. Information about the population of nucleons in HLSPCs in the ground or low-lying excited states relates to some important questions in nuclear physics. For instance, the population in such high-lying orbitals may be caused by tensor forces or nucleon-nucleon correlations \cite{Ong-PLB-2013, Barbieri-PRL-2009}, which are not adequately treated in most traditional shell model calculations \cite{Pandharipande-RMP-1997, Dickhoff-PPNP-2004, Barbieri-PRL-2009}. Such problems may also be linked to the long-disputed questions about the quenching of single particle strength and their dependence on the neutron proton asymmetry, which are suggested to be related to the nucleon-nucleon correlations in nuclear physics \cite{Gade-PRC-2008, Jenny-PRL-2010, Natasha-PRL-2009, Tostevin-PRC-2014, Atar-PRL-2018}.
    
Experimentally, the admixture of high-lying single particle orbitals (SPOs) in the ground states of atomic nuclei can be studied by measuring their charge form factors via $(e,e')$ reactions \cite{Horikawa-PTP-1972,Suzuki-NPA-1977}. However, a hadronic probe is usually more practical in the study of nuclear structure, especially for radioactive nuclei. Actually, in nucleon pickup reactions, \textit{e.g.}, $A(p,d)B$ or $A(\nuc{3}{He},\alpha)B$ reactions, transitions to states of the residual $B$ that are \textit{forbidden} as direct one-step transfer (OST) processes, which would otherwise require picking up neutrons from orbitals that are outside of the model space of traditional shell model, are often observed. These are called ``$j$-forbidden'' transitions \cite{Dehnhard-PR-1967, Fleming-PRev-1968, Nelson-PRC-1972, Clarke-JPG-1979, Kubono-PhD-thesis}. In most of the previous analyses of such reactions, neutron transfers through collective excitations in either the entrance or the exit channels (two-step transfer processes, TST) are thought to be the most important mechanism \cite{Clarke-JPG-1979, Nelson-PRC-1972, Dehnhard-PR-1967, Igarashi-PLB-1972, Abdallah-PRC-1973, Takai-JPSJ-1977, Ishimatsu-NPA-1980, Ohnuma-JPSJ-1980, Kubono-PhD-thesis}. Very few attempts have been made to include OST processes in the analysis of the $j$-forbidden transition data. In Refs. \cite{Dehnhard-PR-1967, Joyce-NPA-1969}, the authors found that including the OST processes only in their calculations using the distorted wave Born approximation (DWBA) failed to reproduce the experimental data with the $j$-forbidden states in both the shapes and the magnitudes of their angular distributions of cross sections. In Refs. \cite{Nelson-PRC-1972} and \cite{Abdallah-PRC-1973}, the authors included both OST and TST processes in coupled reaction channels (CRC) calculations. However, these calculations were used only to examine the neutron spectroscopic amplitudes (SAs) obtained with the Nilsson model by confronting the theoretical results with experimental ones. Extracting the neutron SAs from experimental data was not attempted, which would require a better treatment of the reaction mechanisms and careful parameter searching. In the CRC framework, both OST and TST processes contribute to neutron transfer and they add coherently.

Inclusion of HLSPCs in shell model calculations requires treatment of core excitation and interactions that cross at least two major shells. Neither of these are well investigated. For this reason, HLSPCs are seldom included in standard shell model calculations even with modern computational resources although nucleon population in such orbitals have been suggested a long time ago \cite{Dehnhard-PR-1967, Horikawa-PTP-1972, Warburton-PLB-1992}. Experimental information about nucleon population in these states, such as their SAs, are crucial for the development of structure theory. For this purpose, we study the feasibility of extracting the SAs associated with neutron populations in HLSPCs from experimental data with CRC calculations. 

\section{Procedures for Numerical Analysis}

\begin{figure}[htbp]
        \centering
        \includegraphics[width=0.5\textwidth]{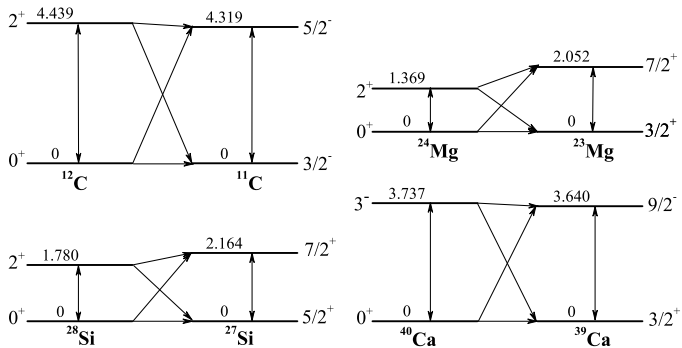}
        \caption{Reaction coupling schemes used in our CRC calculations.}
        \label{fig-01}
    \end{figure}

In order to study the feasibility of extracting the SAs of HLSPCs in the ground and low excited state of nuclei, we reanalyze the $(p,d)$ reactions measured by Ohnuma \textit{et al.} \cite{Ohnuma-JPSJ-1980} at 51.93 MeV with \nuc{12}{C}, \nuc{24}{Mg}, \nuc{28}{Si} and \nuc{40}{Ca} leading to the $j$-forbidden states at 4.32 MeV ($\tfrac{5}{2}^-$) in \nuc{11}{C}, 2.05 MeV ($\tfrac{7}{2}^+$) in \nuc{23}{Mg}, 2.16 MeV ($\tfrac{7}{2}^+$) in \nuc{27}{Si}, and 3.64 MeV ($\tfrac{9}{2}^-$) in \nuc{39}{Ca}. One step transfer to the ground states (g.s.) of the residual and two step transfer to the excited states (e.s.) via collective excitations of the target and of the residual were included in Ref. \cite{Ohnuma-JPSJ-1980}. In addition to these processes, we include direct OST to the $j$-forbidden excited states, assuming the transferred neutrons initially populate the HLSPCs in the ground states of the target nuclei. Neutron pickup from the HLSPCs in the excited states of the target nuclei are also included. For simplicity, only one excited state in the target and the residual nuclei are treated. These processes are depicted in Fig.\ref{fig-01}.

Collective excitations in the entrance- and the exit-channels are treated using a rotational model with deformation parameters taken from Refs. \cite{Ohnuma-JPSJ-1980, B2data-ADNDT-2001, B3data-ADNDT-2002}, which resulted from analysis of the inelastic scattering data. These parameters are fixed in our analysis. The geometry parameters (the radius and diffuseness of a Woods-Saxon potential) of the entrance-channel proton-nucleus potentials are obtained with the systematics of KD02 \cite{kd02}. The depths of these potentials are adjusted to reproduce the elastic scattering data (measured in the same experiment in Ref. \cite{Ohnuma-JPSJ-1980}) in coupled channel calculations. These parameters are given in Table. \ref{tab1}. No deuteron elastic scattering data were measured in the same experiment. Since our main purpose is to study the structure information associated with the $j$-forbidden excited states, to which, contributions from collective excitations of the residual nuclei are very important, we firstly find the depths of the deuteron potentials by requiring them to reproduce the ground state data in single-channel DWBA calculations. Following the prescription in Ref. \cite{Johnson-NPA-1972}, the geometry parameters of these deuteron potentials are taken to be the same as those for the proton-residual potentials. They are also evaluated with the KD02 systematics. The depth of these potentials are then adjusted to reproduce the deuteron elastic scattering cross sections in DWBA calculations by coupled channel calculations. The resulting deuteron potential parameters are also listed in Table. \ref{tab1}. The single-particle form factors are evaluated with the standard binding energy prescription using binding potentials of Woods-Saxon form. We fix the diffuseness parameter $a=0.65$ fm and adopt two values of the radius parameter, namely, $r_0=1.25$ fm and 1.35 fm. We study the dependence of our results on the choice of these values. For convenience, the same $r_0$ is used for all bound state wave functions in our analysis although some studies have suggested to use different $r_0$ values for different bound states (see, e.g., Ref. \cite{Kubono-PhD-thesis}). A systematic study of the proper value of $r_0$ should be further performed in the future. The validity of the following procedures to determine the SAs is rather independent of this problem.
For simplicity, spin-orbit parts of the nuclear potentials are not included in our analysis.
    
    \begin{table}[htbp]
        \caption{Parameters of the entrance- and exit-channel optical model potentials of Woods-Saxon forms: $U(r)=-(V+iW)f_\textrm{ws}(r) - iW_\textrm{d} f_\textrm{wd}(r)$, with $f_\textrm{ws}(r)=\left\{1+\exp\left[\left(r-r_\textrm{V}A_\textrm{T}^{1/3}\right)/a_\textrm{V}\right]\right\}^{-1}$, $f_\textrm{wd}(r)=(4/a)\exp\left[\left(r-r_\textrm{wd}A_\textrm{T}^{1/3}\right)/a_\textrm{wd}\right]\left[1+\exp\left(r-r_\textrm{wd}A_\textrm{T}^{1/3}\right)/a_\textrm{wd}\right]^{-2}$, and $A_\textrm{T}$ being the mass number of the target nucleus. Values of $r_\textrm{v}$, $a_\textrm{v}$, $r_\textrm{wd}$, and $a_\textrm{wd}$ are determined with the systematics of KD02.}
        \begin{center}
                \begin{tabular}{llll}
                    \hline
                    Channel & $V$ (MeV) &  $W$ (MeV) &  $W_\textrm{d}$ (MeV) \\ \hline
                    $p$+\nuc{12}{C}  & 39.26 & 6.795 & 4.377   \\
                    $p$+\nuc{24}{Mg} & 40.22 & 7.457 & 4.355   \\
                    $p$+\nuc{28}{Si} & 39.78 & 6.409 & 4.160   \\
                    $p$+\nuc{40}{Ca} & 40.34 & 6.703 & 4.152   \\
                    $d$+\nuc{11}{C}  & 102.7 & 3.392 & 15.45   \\
                    $d$+\nuc{23}{Mg} & 81.33 & 3.454 & 20.92   \\
                    $d$+\nuc{27}{Si} & 87.02 & 3.342 & 21.40   \\
                    $d$+\nuc{39}{Ca} & 85.80 & 3.404 & 24.66   \\
                    \hline
                \end{tabular}
        \end{center}
        \label{tab1}
    \end{table}

For the excited states of the target nuclei, which have non-zero spins, there are several possible SPOs from which the neutron can be picked up. For example, a neutron is allowed to be in the $1d_{5/2}$, $1d_{3/2}$, $1g_{9/2}$, $1g_{7/2}$, $2d_{3/2}$, $2d_{5/2}$ orbitals (and many other even higher lying ones) when it is picked up from the excited state of \nuc{40}{Ca}, which has spin-parity $3^-$, leaving \nuc{39}{Ca} in its $j$-forbidden excited state, which has spin-parity $\tfrac{9}{2}^-$. For practical reasons, we take only two of them: a ``normal'' one ($1d_{5/2}$ here) and a high-lying one with angular momentum differ from that of the normal one by two units ($1g_{9/2}$ here, governed by requirement of parity conservation). Our choice of these SPOs are listed in Table. \ref{tab2}.

\begin{figure}[htbp]
    \centering
    \includegraphics[width=0.5\textwidth]{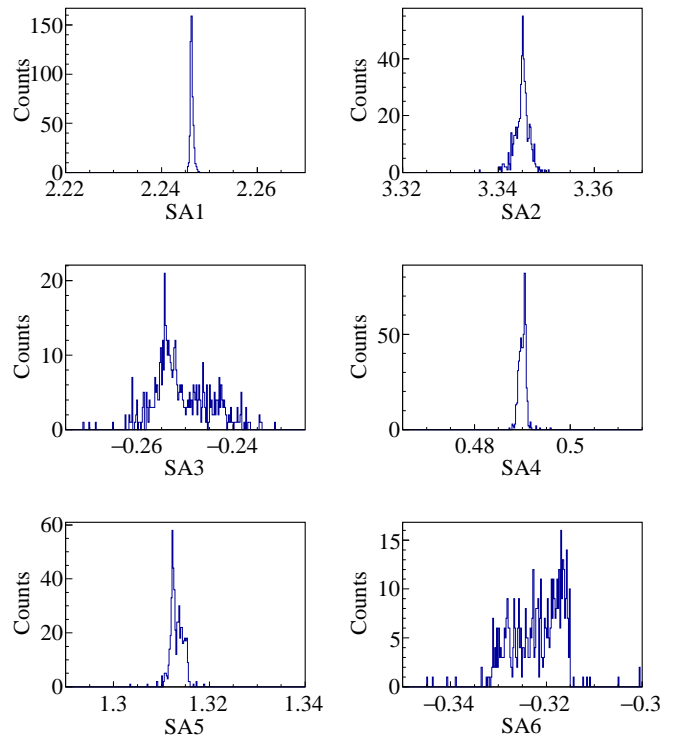}
    \caption{Distributions of the SAs of group G2 of \nuc{40}{Ca}. The total number of parameter set in this group is 515. The mean values, $\overline{\textrm{SA}}$, and their corresponding standard deviations, $\Delta\textrm{SA}$, of these parameters are ($\overline{\textrm{SA}}$, $\Delta\textrm{SA}$)=(2.246, 3.30$\times10^{-4}$) for SA1, (3.345, 1.53$\times10^{-3}$) for SA2, (-0.251, 6.03$\times10^{-3}$) for SA3, (0.490, 7.69$\times10^{-4}$) for SA4, (1.313, 1.38$\times10^{-3}$) for SA5, and (-0.322, 5.39$\times10^{-3}$) for SA6. The single particle orbitals are $1d_{3/2}$ for SA1, $2p_{3/2}$ for SA2, $1f_{7/2}$ for SA3, $1h_{9/2}$ for SA4, $1d_{3/2}$ for SA5, and $1g_{g/2}$ for SA6.}
    \label{fig-02}
\end{figure}

Given the choice of reaction coupling schemes and the involved SPOs, there are six SAs to be determined for each reaction. For a reaction $A(p,d)B$, they are 1) SA1 for OST from the g.s. of $A$ to the g.s. of $B$; 2) SA2 and SA3 for TST from the e.s. of $A$ to the g.s. of $B$; 3) SA4 for OST from the g.s. of $A$ to the e.s. of $B$, and 4) SA5 and SA6 for TST from the e.s. of $A$ to the e.s. of $B$. These SAs are treated as free parameters and their values are determined by fitting the ground and the excited state data simultaneously with the standard minimum $\chi^2$ criteria using the computer code \textsc{SFRESCO} \cite{fresco}. In order to avoid the dependence of the resulting parameters on their initial values, for each reaction, we make more than 1000 times of fittings with the initial SAs randomly chosen within the interval $[-5,5]$. Since for neutron transfer to both the ground and the excited states of residual nuclei there is a set of three SAs to determine, namely, \{SA1, SA2, SA3\} and \{SA4, SA5, SA6\}, respectively, and the values of these two sets of SAs correlate through collective excitations of the target and residual nuclei, one would expect that these SAs may not be able to be uniquely determined by using only two sets of experimental data. Indeed, we have found at least two groups of SAs that fit the experimental data with similar $\chi^2$ values for each reaction. However, it is also interesting to see that the results of more than 1000 times of parameter searches do \textit{gather} into very limited number of groups ($2\sim4$) for each value of $r_0$ ($r_0=1.25$ and 1.35 fm). This reminds us of the discrete ambiguities found in the parameters of systematic optical model potentials \cite{Shkolnik-PLB-1978, Satchler-book-1983}. The relative standard deviations of most SAs, which is defined as the standard deviation of a SA divided with by its mean value, are smaller than 1\% except for SA3, SA4, or SA6 in some groups (corresponding to neutron pickup from HLSPCs), whose relative standard deviations are around 3\%. As an example, we show in Fig. \ref{fig-02} the distributions of 515 groups of SAs which belong to the group G2 of \nuc{40}{Ca}. In the following text, we represent the SAs of each group by their mean values.
    
\section{Results and Discussions} 
   
\begin{figure}[htbp]
    \centering
    \includegraphics[width=0.5\textwidth]{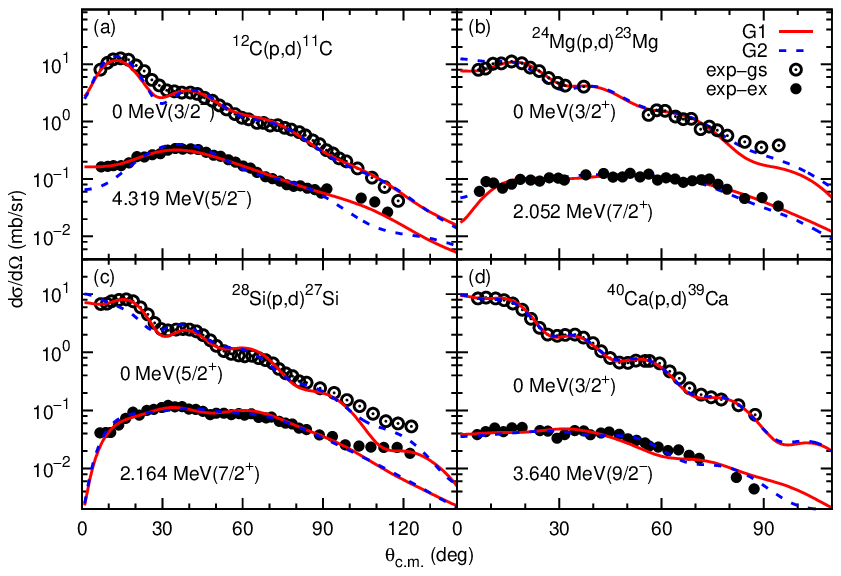}
    \caption{Comparisons between results of CRC calculations and the experimental data with SAs of G1 and G2. Circles and solid points are for the ground and excited state data, respectively. The ground state data of \nuc{24}{Mg} is multiplied by 10 for better visualization.}
    \label{fig-03}
\end{figure}

For each reaction, we present two groups of SAs that give the smallest $\chi^2$ values, which are denoted in Table. \ref{tab2} as groups G1 and G2 for $r_0=1.25$ fm, and G3 and G4 for $r_0=1.35$ fm. Comparisons between the results of CRC calculations with SAs in G1 and G2 are given in Fig. \ref{fig-03} together with the experimental data. Clearly, except for the e.s. data of \nuc{12}{C} at both small and large angles and the g.s. data of \nuc{28}{Si} at forward angles where the SAs of G2 do not reproduce the experimental data as well as those of G1 (although they give very similar $\chi^2$ values), calculations with both groups of SAs reproduced the experimental data nearly equally well. These calculations considerably improved the reproduction of the experimental data as compared with those reported in the original paper \cite{Ohnuma-JPSJ-1980}. Values of spectroscopic amplitudes from available shell model calculations are also given in Table.\ref{tab2} for comparisons with the experimental ones. These calculations were performed with the YSOX \cite{Yuan-PRC-2012} interaction for \nuc{12}{C} and with SPDF-MU \cite{HanRui-spdfmu} for \nuc{24}{Mg}, \nuc{28}{Si} and \nuc{40}{Ca}.

\begin{center}
    \begin{table*}
        \caption{\label{tab2} Mean values of the experimental SAs (SA-EXP) in different groups. The numbers in parentheses are the corresponding $\chi^2$-values. $n,l,j$ are the quantum numbers of the single particle orbitals. Available shell model results are in column SA-SM. See details in the text.}
        \centering
        \begin{tabular}{@{\extracolsep{\fill}}lllllllll}
            \hline
            \multicolumn{2}{c}{nuclei} & $nlj$ &\multicolumn{5}{c}{SA-EXP}   & SA-SM \\\cline{4-9}
            \hline
            \multicolumn{2}{c}{$j^\pi$}& & &\multicolumn{2}{c}{$r_0=1.25$ fm} &\multicolumn{2}{c}{$r_0=1.35$ fm} & \\\cline{1-2}\cline{5-6}\cline{7-8}
            \nuc{12}{C} & \nuc{11}{C} & & &G1 (1.6) &G2 (2.7) & G3 (1.9) & G4 (2.2) &\\\cline{5-6}\cline{7-8}
            $0^{+}$ & ${\frac{3}{2}}^-$ & $1p_{3/2}$ &SA1 & 2.04   &2.16    &1.69    &1.76    & 1.74 \\
            $2^{+}$ & ${\frac{3}{2}}^-$ & $1p_{1/2}$ &SA2 & 3.61   &1.47    &1.56    &-0.210  & 0.740 \\
            &                        & $1f_{7/2}$ &SA3 & 0.708  &0.509   &0.349   &0.170   & \\                                    
            $0^{+}$ & ${\frac{5}{2}}^-$ & $1f_{5/2}$ &SA4 &-0.143  &-0.0928 &-0.0913 &-0.0217 & \\
            $2^{+}$  & ${\frac{5}{2}}^-$ & $1p_{3/2}$ &SA5 & -3.04  &2.52    &-2.51   &2.14    & 1.17\\
            &                        & $1f_{7/2}$ &SA6 & 0.333  &-0.630  &0.0684  &-0.573  & \\\hline                     
            \nuc{24}{Mg} & \nuc{23}{Mg} & & &G1 (1.2) &G2 (1.1) & G3 (1.1)&G4 (0.77) & \\\cline{5-6}\cline{7-8}
            $0^{+}$ & ${\frac{3}{2}}^+$ & $1d_{3/2}$ &SA1 & 0.739 & 0.729  & 0.621  &0.316   & 0.597\\
            $2^{+}$  & ${\frac{3}{2}}^+$ & $1d_{5/2}$ &SA2 &-0.739 & -1.21  &-1.29   &-1.94   & 0.833\\
            &                         & $2s_{1/2}$ &SA3 & 0.881 &-0.643  &-0.416  & -1.25  & 0.138\\                             
            $0^{+}$  & ${\frac{7}{2}}^+$ & $1g_{7/2}$ &SA4 & 0.284 & -0.200 & -0.121 &-0.0218 & \\
            $2^{+}$  & ${\frac{7}{2}}^+$ & $1d_{5/2}$ &SA5 &-1.48  &-0.878  &-0.918  &1.61    & 0.621 \\
            &                         & $1g_{9/2}$ &SA6 &-0.896 & -0.807 &-0.756  & 0.217  & \\\hline                     
            \nuc{28}{Si} & \nuc{27}{Si} & & &G1 (2.9) &G2 (2.9) & G3 (2.3)& G4 (2.3) & \\\cline{5-6}\cline{7-8}
            $0^{+}$  & ${\frac{5}{2}}^+$ & $1d_{5/2}$ &SA1 & 2.39  & 1.95    &2.04    &1.87    & 1.95 \\
            $2^{+}$  & ${\frac{5}{2}}^+$ & $1d_{5/2}$ &SA2 &-1.94  &-2.89    &-1.90   &-2.15   & 0.296\\
            &                         & $2s_{1/2}$ &SA3 & 2.41  &-3.61    & 1.58    & -1.95  & 0.654\\                                    
            $0^{+}$ & ${\frac{7}{2}}^+$ & $1g_{7/2}$ &SA4 & 0.294 &-0.124   & -0.171 &-0.0562 & \\
            $2^{+}$  & ${\frac{7}{2}}^+$ & $1d_{5/2}$ &SA5 & 2.07  & 1.95    &1.80    &1.73    & 1.11 \\
            &                         & $1g_{9/2}$ &SA6 & 0.794 & 0.956   &0.730   &0.746   & \\\hline                           
            \nuc{40}{Ca} & \nuc{39}{Ca} & & &G1 (0.84) &G2 (0.60) & G3 (0.96)& G4 (1.0) & \\\cline{5-6}\cline{7-8}
            $0^{+}$ &  ${\frac{3}{2}}^+$  & $1d_{3/2}$ &SA1 & 2.24   & 2.25  & 1.81   & 1.85    & 2.01 \\
            $3^{-}$ &  ${\frac{3}{2}}^+$  & $2p_{3/2}$ &SA2 & 3.18   & 3.35  & 2.28  & 1.99   & 0.183 \\
            &                         & $1f_{7/2}$ &SA3 &-0.954  &-0.251 & 0.847   & 0.592  & 0.343\\                               
            $0^{+}$ & ${\frac{9}{2}}^-$  & $1h_{9/2}$ &SA4 & 0.337  & 0.490 & 0.460  & 0.289  & \\
            $3^{-}$ & ${\frac{9}{2}}^-$  & $1d_{3/2}$ &SA5 & -0.403 &1.31   & 0.840  &-0.518  & 0.817\\
            &                         & $1g_{9/2}$ &SA6 & -1.21  &-0.322 & -0.264 &-0.823  &\\
            \hline
        \end{tabular}
    \end{table*}
\end{center}

One observes that values of the two groups of SAs with the same $r_0$ differ considerably from each other. Since both sets of SAs result in similar reproduction of the same experimental data, to some extent, the increase/decrease of some SAs, say, SA5, can be compensated by the decrease/increase of the other SAs, say, SA6. This is reminiscent of the continuous ambiguities found in systematic optical model potentials \cite{Satchler-book-1983}. This suggests that some serious problems exist in determining the SAs from experimental data with methods of our kind, which is, however, a rather common practice in the analysis of transfer reactions. On the other hand, since these SAs belong to different groups, if one or some of the SAs in one group can be verified by results of, for instance, nuclear structure calculations or any other means, the SAs of the other groups can be excluded and all the SAs may then be determined.

As stated above, several SPOs may participate when neutron is picked up from the excited states of the target nuclei. For simplicity, we include only two of them. Also, only two excited states of target and residual nuclei are included in our analysis. The SAs extracted from experimental data will change when more SPOs and more excited states are included in CRC calculations. There are also uncertainties in the signs of these SAs. Changing the signs of all SAs simultaneously will result in the same differential cross sections with fixed values of deformation parameters for collective excitation processes. In our approach, the signs of SA1s are set to be positive. The signs of the other SAs are determined by fitting the experimental data. 

Concerning the SAs associated with the HLSPCs in the ground (SA4s) and excited states (SA6s) of the target nuclei, which are the primary interest in this study, one sees that \nuc{40}{Ca} has the most stable (with respect to the change of $r_0$) and largest SA4 value among the four target nuclei. Amplitudes of SA4s for the other nuclei vary much more strongly with respect to the change of $r_0$. This is a strong indication that there is considerable amount of HLSPC in the ground state of \nuc{40}{Ca}. This may be understood by considering that, being in a doubly-magic nucleus, nucleons in \nuc{40}{Ca} have stronger correlations than those in the other three nuclei so that they have larger possibility to populate the high-lying SPOs \cite{Uozumi-NPA-1994}. Although it has been rather well known that there is considerable spread of the single-particle strength in \nuc{40}{Ca} \cite{Mahaux-PR-1985}, we believe this is the first time that we get the SA of the HLSPCs in the ground and excited states of \nuc{40}{Ca} with $(p,d)$ reactions. The amplitudes of SA6 are rather large in most cases for all nuclei, indicating quite large amount of HLSPCs exist in the low-lying excited states of light even-even nuclei.

\begin{figure}[htbp]
    \centering
    \includegraphics[width=0.5\textwidth]{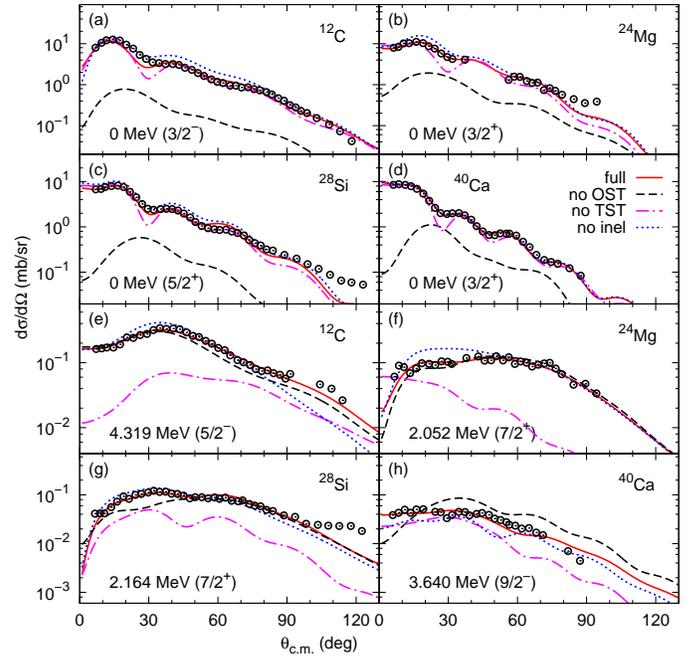}
    \caption{Effects of OST, TST and inelastic excitation processes in neutron pickup reactions. Figures are (a), (b), (c), (d) for the ground state data and (e), (f), (g), (h) for the excited state data of \nuc{12}{C}, \nuc{24}{Mg}, \nuc{28}{Si}, and \nuc{40}{Ca}, respectively. The solid, dashed, dash-dotted, and dotted curves are for results with all processes, without direct OST, without TST via the excited states of the target, and without inelastic excitations in the exit channels, respectively.}
    \label{fig-04}
\end{figure}
        
We further examine how much these different processes contribute to the transfer cross sections. In Fig. \ref{fig-04}, comparisons are made between results of full calculations, which include all processes described previously, with those calculated excluding OST, or TST through the excited states of the target nuclei, or inelastic excitations in the exit channels while keeping all other processes unchanged. Clearly, the OST process is dominant for the ground state data although the TST and inelastic excitations also help to improve the description of the experimental data. On the other hand, the TST processes are dominant for the $j$-forbidden state data. This agrees with the results of previous studies where rough agreements with experimental data were obtained with only TST processes treated for the transfer to the excited states \cite{Ohnuma-JPSJ-1980}. But the contributions from OST processes are also visible, especially for \nuc{28}{Si} and \nuc{40}{Ca}, for which the OST contribute considerably at forward and at all angles, respectively.

\begin{figure}[htbp]
    \centering
    \includegraphics[width=0.5\textwidth]{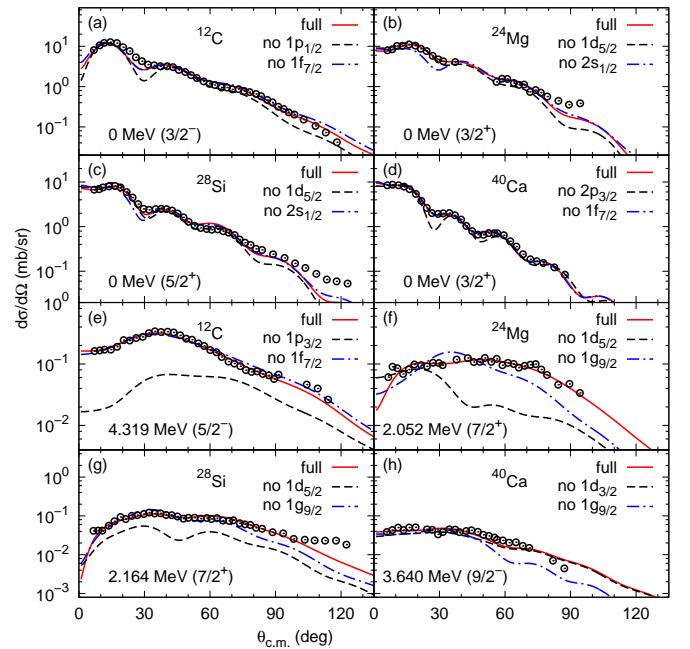}
    \caption{The same as Fig. \ref{fig-04} except that the dashed and dotted curves are obtained by removing one of the two components in neutron pickup from the excited states of the target nuclei leaving the residuals in their ground and excited states while keeping all other reaction processes unchanged.}
    \label{fig-05}
\end{figure}
    
The population of HLSPCs in the excited states of the target nuclei can be examined similarly. For neutron pickup from the excited states of target nuclei, contributions of two components with angular momentum transfer differing by two units are included. Their corresponding SAs are SA2 and SA3 for the ground state data, and SA5 and SA6 for the excited state data. For the excited state data of \nuc{24}{Mg}, \nuc{28}{Si} and \nuc{40}{Ca} and the ground and excited state data of \nuc{12}{C}, one of these components is ``$j$-forbidden''. The effects of these components are examined in Fig. \ref{fig-05} by removing them from the full calculations while keeping all other processes unchanged. One sees that the contributions from these $j$-forbidden components are small for \nuc{12}{C} but are apparent for \nuc{24}{Mg}, \nuc{28}{Si} and \nuc{40}{Ca}.


It is expected that the SAs can be determined uniquely by making more complete calculations, such as including more excited state data in the analysis. As a preliminary test, we analyze the \nuc{24}{Mg}($p,d$)\nuc{23}{Mg} reaction by including three states of the residual (the ground state and the two excited states at 0.45 MeV [$\tfrac{5}{2}^+$] and 2.05 MeV [$\tfrac{7}{2}^+$]). We now have 9 SAs as free parameters to determine. Again, our calculations show that we have two groups of SAs that give similar $\chi^2$ values but rather different SAs. These results are given in Table. \ref{tab-6p-9p} together with the results of the 6-parameter fitting as shown in Table. \ref{tab2}. The other three SAs which correspond to nucleon transfer to the excited state of \nuc{23}{Mg} at 0.45 MeV are not shown. One observes that the SAs of the group 9P-Set1 are rather close to their counterparts in the group 6P-G2 and their signs are all in consistent. On the other hand, the SAs of the group 9P-Set2 are different from both groups of 6-parameter searching. This turns out to be an excellent example showing that the inclusion of more excited state data may better constrain the SAs. From this result, we may exclude the parameters in the group G1 in Table.\ref{tab2} and update the SAs of group G2 by those of 9P-Set1 in Table. \ref{tab-6p-9p} for the SAs of the \nuc{24}{Mg} case.

\begin{table}[htbp]
    \centering
    \caption{SA$i$ ($i=1,\cdots,6$) from 6-parameter and 9-parameter searching for the \nuc{24}{Mg}($p,d$)\nuc{23}{Mg} reaction. See text for details. The single particle orbitals corresponding to these SAs are indicated in the second row.}
    \begin{tabular}{crrrrrr}\hline
        & \multicolumn{1}{c}{SA1} & \multicolumn{1}{c}{SA2} & \multicolumn{1}{c}{SA3} & \multicolumn{1}{c}{SA4} & \multicolumn{1}{c}{SA5} & \multicolumn{1}{c}{SA6} \\
        & \multicolumn{1}{c}{$1d_{3/2}$} & \multicolumn{1}{c}{$1d_{5/2}$} & \multicolumn{1}{c}{$2s_{1/2}$} & \multicolumn{1}{c}{$1g_{7/2}$} & \multicolumn{1}{c}{$1d_{5/2}$} & \multicolumn{1}{c}{$1g_{9/2}$} \\\hline
        6P-G1 & 0.739  & -0.739  & 0.881  & 0.284  & -1.480  & -0.896  \\
        6P-G2 & 0.729  & -1.210  & -0.643  & -0.200  & -0.878  & -0.807  \\
        9P-Set1 & 0.659  & -1.124  & -0.437  & -0.334  & -0.666  & -0.887  \\
        9P-Set2 & 0.432  & -2.000  & -1.225  & -0.255  & 1.283  & 1.838  \\\hline
    \end{tabular}%
    \label{tab-6p-9p}%
\end{table}%

\section{Summary}

In Summary, $(p,d)$ reactions with \nuc{12}{C}, \nuc{24}{Mg}, \nuc{28}{Si}, and \nuc{40}{Ca} at an incident energy of 51.93 MeV leaving the residual nuclei in their ground and $j$-forbidden excited states are analyzed with CRC calculations. SAs for neutrons in normal and HLSPCs in the ground and excited states of these even-even light nuclei are obtained by simultaneously fitting the ground state and the excited state data. Continuous and discrete ambiguities are seen in these SAs, which are similar to the ambiguities found in systematic optical model potentials. Although made with limited number of reaction channels and the resulting SAs have uncertainties from several sources, our results show that they can be confined to within very few discrete groups. Our preliminary tests with the \nuc{24}{Mg}(p,d)\nuc{23}{Mg} reaction suggests that it is possible to further determine these SAs uniquely by including data of more excited states in the analysis. The present analysis demonstrates for the first time that the information of HLSPCs can be obtained with nucleon pickup reactions. A combined efforts on measurements of (p,d) reactions and reaction model analysis, such as the one outlined in this paper, for the experimental SAs of the HLSPCs and development of future structure theories, which may treat these components adequately, will certainly help us to understand some important problems in nuclear physics, such as the nucleon-nucleon correlations in atomic nuclei.

\begin{acknowledgments}
This work is supported by the National Natural Science Foundation of China (Grants Nos.  U1432247, 11775013, and 11775316) and the national key research and development program (2016YFA0400502). C.A.B. acknowledges support by the U.S. DOE grant DE-FG02-08ER41533, the U.S. National Science Foundation Grant No. 1415656, the U.S. DOE grant DE-FG02-13ER42025, and the China-U.S. Theory Institute for Physics with Exotic Nuclei (CUSTPEN).
\end{acknowledgments}

\bibliographystyle{apsrev4-1}
\bibliography{pd-hlspcs-2018}

\end{document}